# A simple Urea approach to N-doped α-Mo$_2$C with enhanced superconductivity

Longfu Li(李龙夫)[1], Lei Shi(石磊)[2], Lingyong Zeng(曾令勇)[1], Kuan Li(李宽)[1], Peifeng Yu(余沛峰)[1], Kangwang Wang(王康旺)[1], Chao Zhang(张超)[1], Rui Chen(陈睿)[1], Zaichen Xiang(项载琛)[1], Yunwei Zhang(张云蔚)[2], Huixia Luo(罗惠霞)[1,*]

[1]School of Materials Science and Engineering, State Key Laboratory of Optoelectronic Materials and Technologies, Guangdong Provincial Key Laboratory of Magnetoelectric Physics and Devices, Key Lab of Polymer Composite & Functional Materials, Sun Yat-Sen University, Guangzhou 510275, China

[2]School of Physics, Guangdong Provincial Key Laboratory of Magnetoelectric Physics and Devices, Sun Yat-Sen University, 510275 Guangzhou, China

China

[*]Corresponding author. Email: *luohx7@mail.sysu.edu.cn* (H. Luo)

**Abstract:** Chemical doping is a critical factor in the development of new superconductors or optimizing the superconducting transition temperature ($T_c$) of the parent superconducting materials. Herein, a new simple urea approach is developed to synthesize the N-doped α-Mo$_2$C. Benefiting from the simple urea method, a broad superconducting dome is found in the Mo$_2$C$_{1-x}$N$_x$ ($0 \leq x \leq 0.49$) compositions. XRD results show that the structure of α-Mo$_2$C remains unchanged and that there is a variation of lattice parameters with nitrogen doping. Resistivity, magnetic susceptibility, and heat capacity measurement results confirm that the superconducting transition temperature ($T_c$) was strongly increased from 2.68 K ($x = 0$) to 7.05 K ($x = 0.49$). First-principles calculations and our analysis indicate that increasing nitrogen doping leads to a rise in the density of states at the Fermi level and doping-induced phonon softening, which enhances electron-phonon coupling. This results in an increase in $T_c$ and a sharp rise in the upper critical field. Our findings provide a promising strategy for fabricating transition metal carbonitrides and provide a material platform for further study of the superconductivity of transition metal carbides.

TMCs (transition metal carbides) are a large class of materials that combine the properties of ceramics and metals due to the incorporation of carbon atoms into their metal lattice. The current high interest in TMCs is due mainly to their unique physical-chemical properties and wide range of potential applications. TMCs exhibit high chemical stability and functional physical properties, such as high hardness, high wear resistance, high conductivity, high melting point, acid and alkali resistance, or even superconductivity [1-7].

Recently, the two-dimensional (2D) TMCs (MXene, with the formula of $M_{n+1}X_nT_x$, where n ranges from 1 to 4; M represents early transition metals (such as Ti, V, Cr, etc.); X represents C or/and N and $T_x$ denotes surface functionalization) can be obtained via selective etching the "A" layers from MAX phases[8,9] (MAX, with the formula of $M_{n+1}AX_n$, A represents an element from group III-A or IV-A of the periodic table (such as Al, Si, etc.) with hydrofluoric acid[10,11]. They have been proven to possess many intriguing properties, such as electrochemical energy storage, efficient thermoelectric conversion, and superconductivity[1,9,10]. However, most of the synthesized MXenes end in functional groups such as hydroxyl and oxygen[9,11], which require some post-treatment for purification. Previous reports have shown that large-area, high-quality 2D TMC crystals can also be successfully grown by chemical vapor deposition (CVD) [14]. $Mo_2C$, one of the most widely studied 2D TMCs, has also been successfully synthesized by CVD methods[15-17]. Previous results indicate that different growth conditions have an impact on the size, quality, crystal structure, and physical properties of the $Mo_2C$ crystals. It is found that the 2D characteristics of superconducting transitions of 2D orthorhombic α-$Mo_2C$ are consistent with Berezinskii-Kosterlitz-Thouless behavior and show strong anisotropy with magnetic field orientation, as well as strong dependence on the crystal thickness. Due to the possible presence of phonon stiffening, the $T_c$ of 2D hexagonal β-$Mo_2C$ decreases monotonically with increasing pressure. In contrast, 2D α-$Mo_2C$ exhibits dome-shaped superconductivity under pressure, indicating the presence of two competing effects arising from phonon and electronic characteristics. From the above discussion, we know that there is still a lack of inexpensive and simple synthesis methods to obtain high-quality 2D $Mo_2C$. This hinders more extensive and in-depth research into its intrinsic physical properties.

The nitrogen-doped strategy is an effective method for optimizing superconductivity[18-20] as electron-rich doping increases the density of states (DOS), which is achieved by increasing the Fermi surface[21]. $Mo_2C$ and $Mo_2N$ have been studied as superconductors for many years[2-4,22-24]. More recently, first-principles electronic structure calculations indicate that the α-$Mo_2C$ superconductor has nontrivial topological electronic band structures[25]. Therefore, α-$Mo_2C$ provides a promising material platform for exploring topological superconductivity and Majorana zero modes. Theoretical calculations show that the topological surface states on the (001) surface of α-$Mo_2C$ intersect the Fermi level. These surface states possess helical spin textures and can induce equivalent $p + ip$ superconductivity through the proximity effect. It remains inconclusive whether the nontrivial topological electronic bands of undoped α-$Mo_2C$ can be retained after N doping. Besides, there is limited research on the effects of nitrogen doping on the superconductivity of transition metal carbides, hindering the understanding of the mechanisms of superconductivity. Therefore, we consider using nitrogen doping to study the alteration of the superconductivity of α-$Mo_2C$. There are several reports of nitrogen-doped $Mo_2C$, but these reports used many complex synthesis methods and reported very limited or inaccurate control of the nitrogen content of the samples [26-29].

Here, we report a new and straightforward urea method for preparing $Mo_2C_{1-x}N_x$ ($0 \leq x \leq 0.49$) samples. The lattice parameters of molybdenum carbide undergo anisotropic changes when carbon is partially replaced by nitrogen. Noticeably, as the nitrogen content increases, the $T_c$ of $Mo_2C_{1-x}N_x$ rises from 2.68 to 7.05 K. The analysis of the heat capacity data illustrates that $Mo_2C_{1-x}N_x$ are bulk superconductors. Density functional theory (DFT) theoretical calculations were also conducted to show the electronic band structure of N-doped α-$Mo_2C$. First-principles calculations demonstrate that N doping leads to an increase of the DOS at the Fermi level, consistent with the increasing $T_c$ observed experimentally.

Molybdenum powder (99.9%, 250 mesh, Alfa Aesar) and carbon (Macklin) are used as raw materials for α-$Mo_2C$. Specifically, 300 mg of the powders with a molar ratio of 2:1 were pressed into a block and then arc-melted under an argon atmosphere. In order to synthesize a series of $Mo_2C_{1-x}N_x$ ($0 \leq x \leq 0.49$), the as-obtained α-$Mo_2C$ was mixed with different stoichiometric urea (99.999%, Aladdin) and ground into powder. Next, the resulting mixed powder was pyrolyzed at 800 °C (heating rate: 5 K min$^{-1}$) under a nitrogen atmosphere for 3 h. Finally, the black powder was pressed into a piece and placed into a quartz tube, which was sealed in a high vacuum ($< 1 \times 10^{-1}$ MPa) and annealed at 1000 °C to obtain a denser powder for subsequent performance testing. The porosity of sintered pellets was determined by Archimedes' method.

The crystal structure information of the prepared samples was tested and analyzed by powder X-ray diffraction (PXRD). The XRD patterns were collected using the MiniFlex of Rigaku (Cu Kα radiation) with a step width of 0.01° and a constant scan speed of 1 °/min from 10 to 90° and a sealed-tube X-ray generator with the Cu target operating at 45 kV and 15 mA. The raw XRD profiles were refined using the Rietveld method in the FullProf suite package software. The surface states and elemental compositions of all samples were measured by X-ray photoelectron spectroscopy (XPS, Nexsa, Thermo Fisher Scientific, xenon lamp as light source, energy resolution < 0.45 eV) in an ultra-high vacuum atmosphere. The as-obtained binding energy data were calibrated against the binding energy of C1s orbital (284.8 eV). The elemental ratio was determined via an Elemental analyzer (Elementar Vario EL cube). Scanning electron microscopy (SEM) combined with energy-dispersive X-ray spectroscopy (SEM-EDS) was used to collect the surface morphology and element distribution of the $Mo_2C_{1-x}N_x$. The relevant electrical transport and specific heat, as well as magnetization properties of as-prepared $Mo_2C_{1-x}N_x$ samples, were measured in detail in the Quantum Design PPMS-14T, and resistivity was measured using a standard four-probe strategy.

The electronic structures of α-$Mo_2C$ and doping different proportions of N element were studied on the basis of the first-principles calculations[30]. The projector-enhanced method[31], as introduced in the VASP package[32,33], was employed to characterize both the core and valence electrons, as well as the core electrons. The generalized gradient approximation (GGA) [34] of the Perdew-Burke-Ernzerhof (PBE) type was chosen for the exchange correlation function. The plane-wave basis was assigned a kinetic energy cutoff of 520 eV. The structures were completely relaxed until the forces on all atoms were less than 0.01 eV/Å.

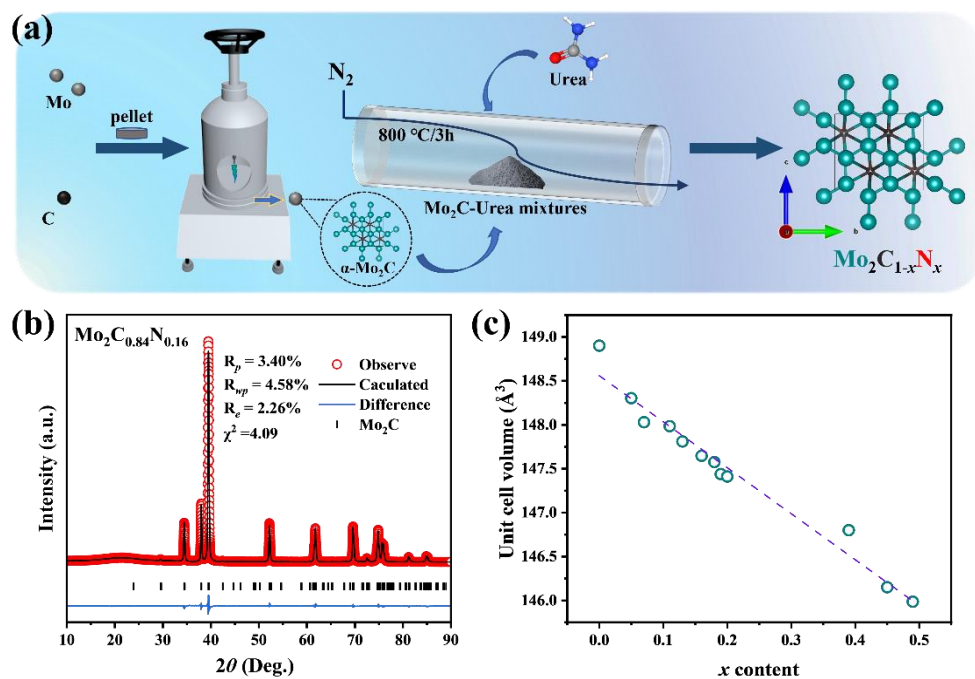

Fig. 1. (a) Two-step synthetic strategy for $Mo_2C_{1-x}N_x$ ($0 \leq x \leq 0.49$) samples. (b) Observed and calculated XRD patterns of $Mo_2C_{0.84}N_{0.16}$ sample. (c) The trends in unit cell volume as the change of the N doping amount.

As shown in Fig.1(a), we have developed a simple methodology for nitrogen doping based on urea. In the first step, molybdenum powder and carbon powder at a molar ratio of 2:1 were pressed into a block and then arc-melted to obtain α-$Mo_2C$. To synthesize $Mo_2C_{1-x}N_x$, we carefully selected urea as our nitrogen source. This is because urea plays a vital role as an "intermediate solvent" and stabilizer during the mixing process[35,36]. The α-$Mo_2C$ synthesized in the first step was ground with urea into a uniform powder and treated at 800 °C for 3 hours in a nitrogen atmosphere at a heating rate of 5 °C/min, followed by natural cooling. The nitrogen flow rate through the reactor was approximately 20 mL/min. During the reaction, C atoms in α-$Mo_2C$ are replaced by N atoms, resulting in nitridation, and $Mo_2C_{1-x}N_x$ as black powder was obtained. We can control the amount of nitrogen doping in the sample by adjusting the proportion of urea in the precursor. It can be seen that this method is solvent-free and has a facile post-treatment.

Figures 1 and S1 display the PXRD patterns of the $Mo_2C_{1-x}N_x$ ($0 \leq x \leq 0.49$) samples and the results of the associated data analysis. The samples were analyzed for carbon and nitrogen content using an Elemental Analyzer. By this method, we obtained a maximum doping of 0.49 (see Table S1). A representative $Mo_2C_{0.84}N_{0.16}$ sample was selected for demonstration, and the XRD Rietveld refinement is shown in Fig. 1(b), which can be indexed by orthorhombic α-Mo$_2$C (PDF card number:04-001-2151) with space group *Pbcn*. The XRD Rietveld refinements for other compositions are shown in Fig. S2. The crystal structure of Mo$_2$C$_{1-x}$N$_x$ is in Fig. 1(a). Some samples included small amounts of molybdenum impurities [see Fig. S1(a)]. A low content of Mo will not have a significant effect on its crystal structure and property measurements. The peaks around 38° were amplified in Fig. S1(b). As the amount of nitrogen elemental gradually increased, a leftward shift of the peak can be perceived, representing a shift of the (200) crystal plane to a lower angle. The lattice parameter increases from 4.7308(1) Å of α-Mo$_2$C to 4.7593(5) Å of Mo$_2$C$_{0.51}$N$_{0.49}$. As shown in Figs. 1(c) and S1(b), the lattice parameter *a* increases, while *b* and unit cell volume decrease with increasing nitrogen doping content. Both the lattice parameter *a* and unit cell volume exhibit an almost linear relationship with *x*.

The SEM-EDS characterization of Mo$_2$C$_{0.51}$N$_{0.49}$ and α-Mo$_2$C samples was performed in order to check the homogeneity and observe the surface morphology. As displayed in Fig. S3, all the constituent elements are uniformly distributed. Surprisingly, it was found that it is possible to synthesize layered α-Mo$_2$C by a simple arc-melting method, which is an improvement over the synthesis of layered MXenes by acid etching. Thus, this provides an excellent platform for our subsequent rapid synthesis and investigation of the physical properties of 2D transition metal carbides. In addition, the XPS test was used to investigate the chemical state of the nitrogen element in the Mo$_2$C$_{1-x}$N$_x$ ($0 \leq x \leq 0.49$) samples. In Fig. 2(b), the high-resolution C 1s spectrum can be well-fitted into the Mo-C, C-C, N-C, and O-C=O[37,38], respectively. In Fig. 2(c), the Mo 3d peaks can be deconvolved into six peaks, suggesting that the Mo element exists in the Mo$_2$C$_{0.51}$N$_{0.49}$ sample in three different states[39,40]. Strikingly, the peaks with binding energies of 229.0 and 232.7 eV are attributed to the bimodal Mo—N bond. Moreover, in Fig. 2(d), the peak around 396 eV also corresponds to the Mo—N bond[41]. Apparently, for Mo$_2$C$_{0.51}$N$_{0.49}$, there are more Mo—N bonds than Mo$_2$C$_{0.93}$N$_{0.07}$. The excellent agreement with the analytical results of PXRD spectra further reveals the successful N doping into α-Mo$_2$C.

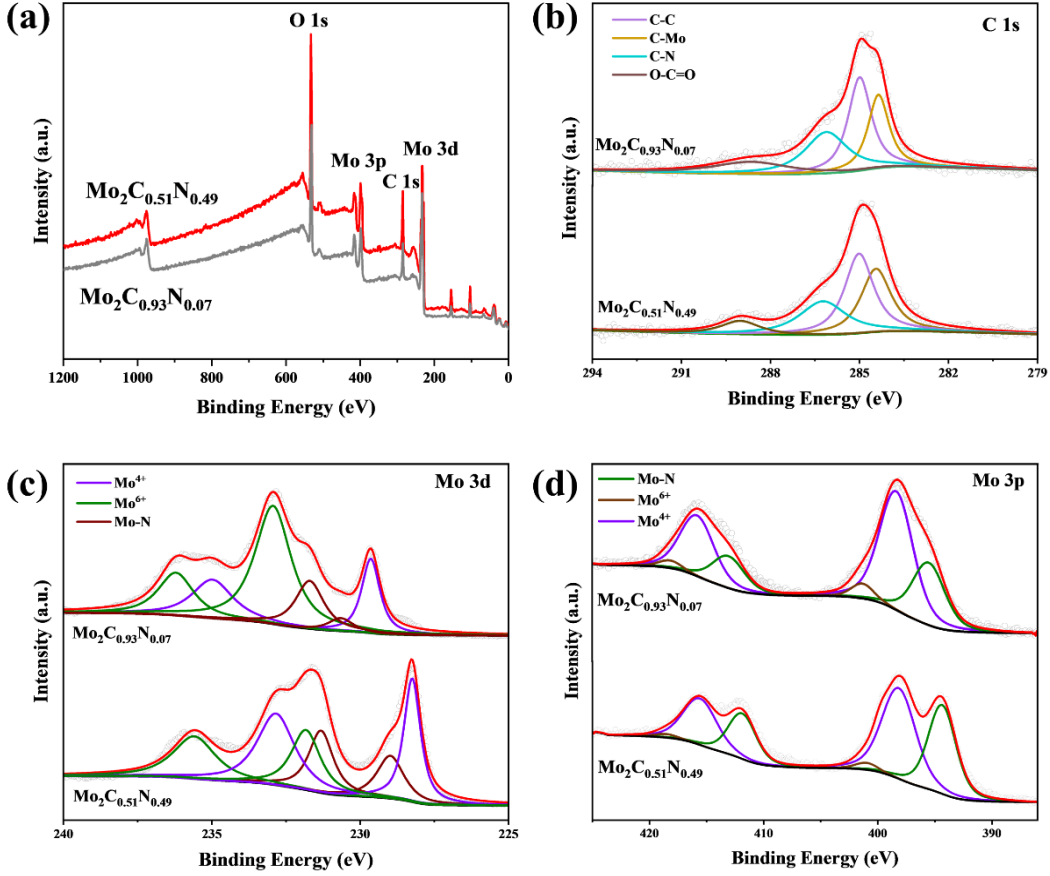

Fig. 2. (a) XPS survey spectra for $Mo_2C_{0.93}N_{0.07}$ and $Mo_2C_{0.51}N_{0.49}$ in (b) C 1s; (c) Mo 3d and (d) Mo 3p regions.

We first performed resistivity measurements to investigate the superconducting properties in the nitrogen-doped α-$Mo_2C$ compounds. Fig. 3(a) shows the resistivity versus temperature from 300 to 1.8 K. The resistivity of all the samples decreased with decreasing temperature and showed metallicity. Fig. 3(b) shows the resistivity transformation for all samples more clearly at low temperatures. Superconductivity manifests as a sharp decrease in electrical resistivity until it drops to zero. We use the criterion that a 50 % fixed percentage of the normal-state resistivity is defined as the $T_c$ of the sample, α-$Mo_2C$ synthesized by us, to be around 2.7 K, which is consistent with previous reports[3]. The $T_c$ of the sample with the maximum doping amount $Mo_2C_{0.51}N_{0.49}$ is 7.05 K, indicating that nitrogen doping has a significant impact on the $T_c$. It is worth noting that the amount of $T_c$ enhanced by our doping method is higher than in a similar report[42]. Further, the resistivity curve of the vast majority of $Mo_2C_{1-x}N_x$ ($0 \leq x \leq 0.49$) samples reveal superconducting transition widths of less than 0.3 K, indicating that the samples synthesized by this method are of high purity. The results suggest that nitrogen doping indeed enhances the superconductivity of α-$Mo_2C$. In addition, we recorded the superconducting transition temperature values and the amount of nitrogen doping $x$ for each sample to establish the electronic phase diagram.

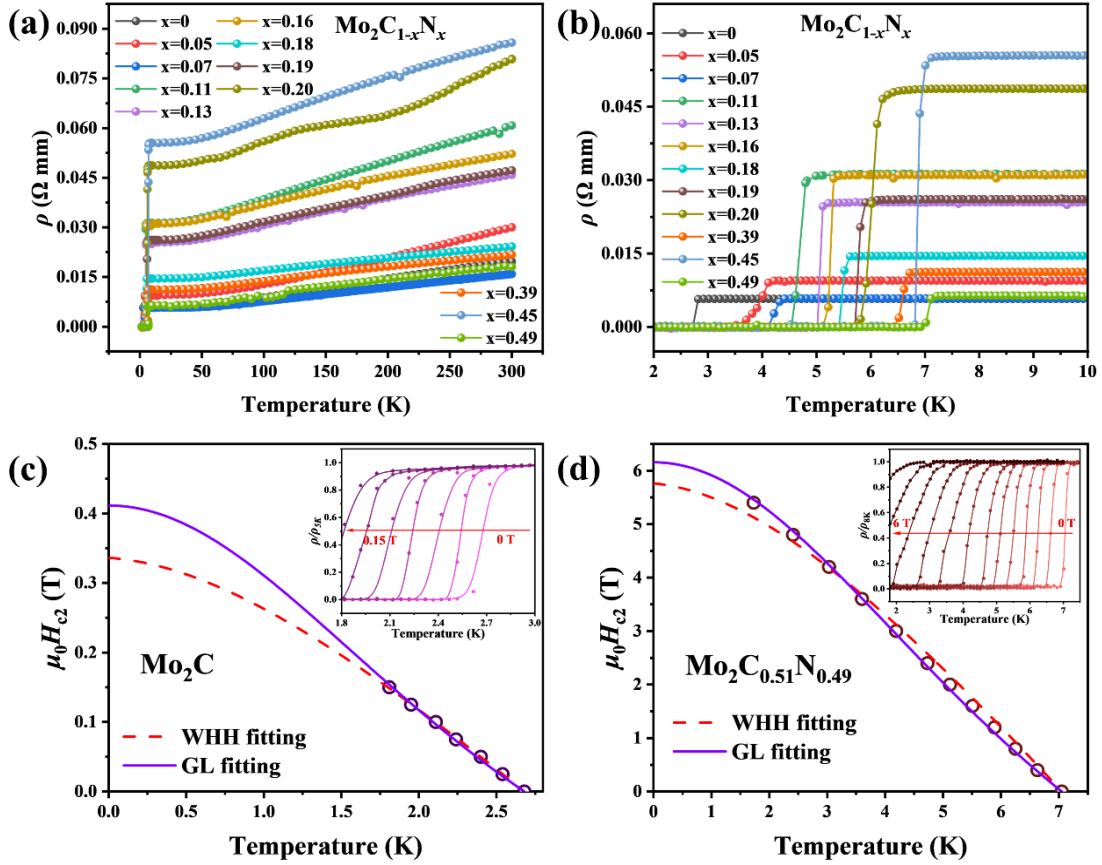

Fig. 3. The measurements of temperature-dependent resistivity and upper critical ($\mu_0 H_{c2}$) fields for $Mo_2C_{1-x}N_x$ ($0 \leq x \leq 0.49$). (a) The temperature dependence of resistivity; (b) The resistivity at low temperatures. (c) and (d) show the $\mu_0 H_{c2}-T$ phase diagram. The violet curve exhibits refinement using the G-L theory, while the red curve demonstrates refinement using the WHH model. The upper right inset shows the resistivity at different applied magnetic fields.

The resistivity under various magnetic fields of $Mo_2C_{1-x}N_x$ ($x = 0$ and 0.49) is emphasized in the inset of Figs. 3(c) and 3(d), which show that $T_c$ decreases with increasing magnetic field, and the superconducting transition gets progressively wider. We used the same criteria as for the zero-field resistivity data to obtain the upper critical field $\mu_0 H_{c2}(0)$ for α-$Mo_2C$ and $Mo_2C_{0.51}N_{0.49}$ samples, displayed in Figs. 3(c) and 3(d). $\mu_0 H_{c2}(0)$ can be obtained from calculations of the Werthamer-Helfand-Hohenburg (WHH) model and Ginzberg-Landau (G-L) theory. Based on the WHH model, we can calculate the $\mu_0 H_{c2}(0)$ for α-$Mo_2C$ and $Mo_2C_{0.51}N_{0.49}$ as 0.33 and 5.75 T, respectively. According to the prerequisites as derived from the Pauli limiting effect[43], the calculated $\mu_0 H_{c2}(0)$ through the WHH model has to be smaller than the Pauli paramagnetic limit $H^P = 1.86*T_c$. The $H^P$ of α-$Mo_2C$ and $Mo_2C_{0.51}N_{0.49}$ is 4.98 T and 13.11 T, respectively, which are higher than those calculated by the WHH model, implying that the orbital effect limits it. The distribution of functions based on G-L theory: $\mu_0 H_{c2}(T) = \mu_0 H_{c2}(0) * [1-(T/T_c)^2]/[1+(T/T_c)^2]$, where $\mu_0 H_{c2}(0)$ can be calculated. As shown in the figure, the data points fit the function well, and the upper critical fields for α-$Mo_2C$ and $Mo_2C_{0.51}N_{0.49}$ are 0.41 and 6.16 T, respectively.

**Table 1.** Comparison of the Superconducting Parameters of $Mo_2C_{1-x}N_x$ ($x = 0$, $x = 0.11$, $x = 0.49$). $\gamma$ is the electronic specific heat constant, $\beta$ is the phonon contribution factor. $\Theta_D$ is Debye temperature. $\Delta C/\gamma T_c$ is the value of the normalized specific heat jump. $\lambda_{ep}$ is the constant of electron-phonon coupling.

| Compound | $Mo_2C$ | $Mo_2C_{0.89}N_{0.11}$ | $Mo_2C_{0.51}N_{0.49}$ |
|---|---|---|---|
| $T_c$ (K) | 2.68 | 4.69 | 7.05 |
| $\mu_0H_{c2}$ (T) (WHH) | 0.33 | 1.53 | 5.75 |
| $\mu_0H_{c2}$ (T) (G-L) | 0.41 | 1.71 | 6.16 |
| $H^P$ (T) | 4.98 | 8.72 | 13.11 |
| $\gamma$ (mJ mol$^{-1}$ K$^{-2}$) | 5.618 | 7.993 | 9.581 |
| $\beta$ (mJ mol$^{-1}$ K$^{-4}$) | 0.089 | 0.155 | 0.108 |
| $\Theta_D$ | 402.8 | 334.9 | 377.7 |
| $\Delta C/\gamma T_c$ | 1.24 | 1.08 | 0.95 |
| $\lambda_{ep}$ | 0.51 | 0.61 | 0.66 |
| $N(E_F)$ (states/eV) | 6.544 | 6.920 | 7.887 |
| $\mu_0H_{c1}$ (Oe) | / | / | 105.7 |
| $\xi_{GL}(0)$ (nm) | 28.3 | 13.8 | 7.2 |

To understand the superconducting state further, the upper critical magnetic fields were systematically investigated for all samples by measuring the change in resistivity of the material with the temperature at different applied magnetic fields. In Fig. S4 and Table S2, we have calculated and summarized some superconducting parameters for all samples for comparison. It can be seen that the GL model fits the experimental data satisfactorily in the entire temperature range, and the $\mu_0H_{c2}(0)$ increases with increasing N content. The value of $\mu_0H_{c2}(0)$ from the WHH model and G-L theory reveal that the $\mu_0H_{c2}(0)$ is obviously enhanced with nitrogen doping. Additionally, we can calculate the GL coherence length $\xi_{GL}(0)$ by the formulation $\mu_0H_{c2}(0) = \Phi_0/2\pi\xi_{GL}^2$, where $\Phi_0$ represents the quantum flux ($h/2e$). The obtained $\xi_{GL}(0)$ decreases from 28.3 nm to 7.2 nm with increasing $x$ from 0 to 0.49. It can be seen that $\mu_0H_{c2}(0)$ is enhanced while the coherence length $\xi_{GL}(0)$ is shortened, suggesting that there is a stronger electron-electron interaction in the nitrogen-doped α-$Mo_2C$[44]. From these facets, we speculate the shorter coherence length may induce the enhanced $\mu_0H_{c2}(0)$ in the nitrogen-doped samples due to the electron scattering[45,46].

Figure 4(a) displays the temperature-dependent magnetizations during the zero field cooling (ZFC) process under an external field of 3mT. Diamagnetic shielding and a clear superconducting transition were observed, confirming the occurrence of bulk superconductivity in $Mo_2C_{1-x}N_x$ ($0 \leq x \leq 0.49$) samples. The $T_c$ is slightly lower than that obtained from the $R$-$T$ curve due to the suppression effect of the external field. Since the prevalence of Meissner screen current decay in polycrystalline samples[47], we will find a broader superconducting transition. In order to further investigate the superconducting properties, we conducted the magnetization measurements at low temperatures for the sample $Mo_2C_{0.51}N_{0.49}$. Fig. 4(b) displays the M-H curves. The value of magnetization increases linearly and decreases after reaching the lower critical field. We can see that it eventually turns into a paramagnetic state, exhibiting a typical type-II superconductor behavior. $M_{fit} = e + f H$ was determined by linearly fitting the low-field region of the magnetization[48]. When the difference between $M$ and $M_{fit}$ exceeds 1% $M_{max}$, the value of $\mu_0 H_{c1}(0)$ can be extracted. As is shown in Fig. S5(b), the location of data point extraction in Fig. S5(a) was indicated by the grey line. The extracted points are fitted according to the formula $\mu_0 H_{c1}(T) = \mu_0 H_{c1}(0)\left[1-(T/T_c)^2\right]$, where $\mu_0 H_{c1}(0)$ is the lower critical field. The value of $\mu_0 H_{c1}(0)$ can be determined to be 105.7 Oe for the $Mo_2C_{0.51}N_{0.49}$ sample.

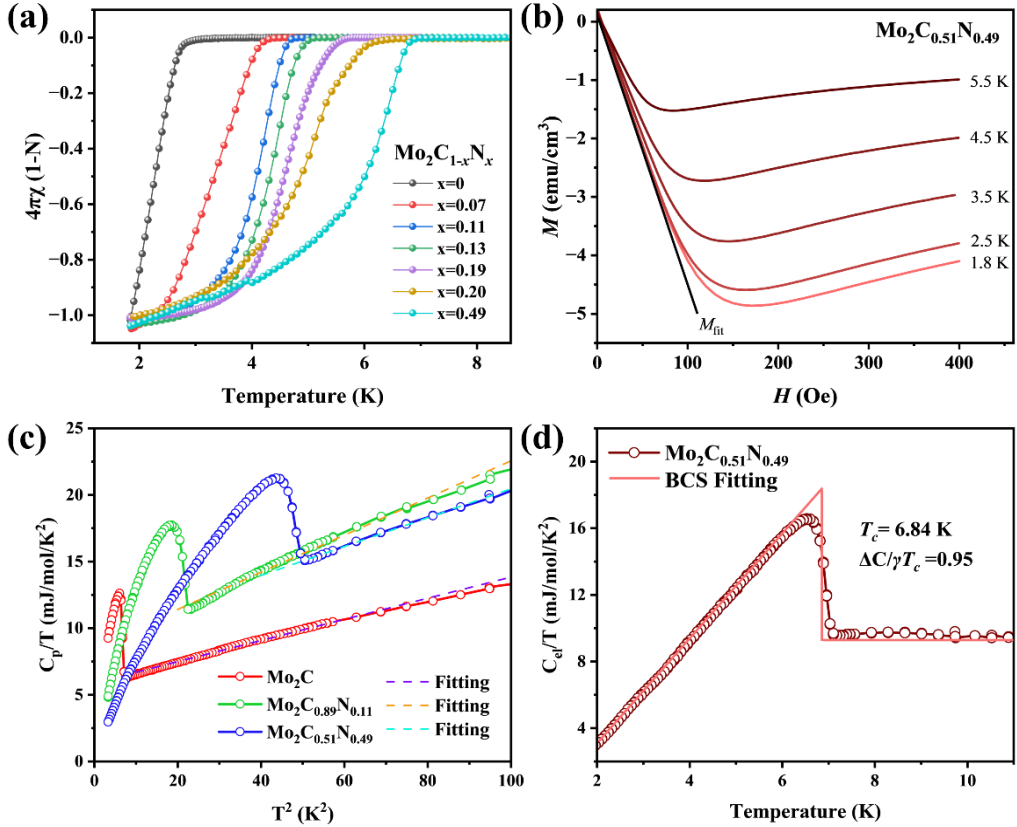

Fig. 4. (a) The ZFC magnetic susceptibility with temperature for $Mo_2C_{1-x}N_x$ ($0 \leq x \leq 0.49$) samples under applied magnetic field 3 mT. (b) The isothermal magnetization curves for $Mo_2C_{0.51}N_{0.49}$. (c) The specific heat data for $Mo_2C_{1-x}N_x$ ($x = 0$, $x = 0.11$, $x = 0.49$), The data are plotted as $C_p/T$ vs. $T^2$ diagram. (d) The temperature dependence of the electronic contribution to the specific heat for $Mo_2C_{0.51}N_{0.49}$.

The transition from the normal state to the superconducting state causes a sudden change in specific heat. For this reason, low-temperature intrinsic property heat capacity measurements were performed under a zero applied field to confirm whether the polycrystalline sample $Mo_2C_{1-x}N_x$ ($x = 0, 0.11, 0.49$) is a bulk superconductor. The heat capacity measurement is analyzed as depicted in Fig. 4(c), and a distinct jump in specific heat is evident, which indicates the bulk nature of superconductivity. The heat capacity data for the normal state can be matched by the formula $C_p = \gamma T + \beta T^3$, where $\beta$ is the specific heat coefficient of the lattice part, and $\gamma$ represents the electronic specific heat coefficients, which are mainly contributed by phonons and electrons, respectively. For the $Mo_2C_{0.51}N_{0.49}$ sample, the values of $\gamma$ and $\beta$ are 9.581 mJ mol$^{-1}$ K$^{-2}$ and 0.108 mJ mol$^{-1}$ K$^{-4}$, respectively. Figure 4(d) displays the curve of $C_{el}/T$ vs. $T$, where $C_{el}$ is given by the expression $C_{el} = C_p - \beta T^3$. Taking the equal-area entropy construction into account, $T_c = 6.84$ K is derived from the electronic-specific calorimetry data and is consistent with the results of the above two tests. Compared to the Bardeen-Cooper-Schrieffer (BCS) weak-coupling limit value of 1.43, the normalized specific heat jump value of $\Delta C/\gamma T_c$ for $Mo_2C_{1-x}N_x$ ($x = 0$, $x = 0.11$, $x = 0.49$) are close and decrease with the increase of nitrogen doping amount. Table 1 shows that the $\Delta C/\gamma T_c$ values for $Mo_2C_{1-x}N_x$ are relatively low, indicating the system's weak coupling strength.

For the sake of contrast, we also measured the heat capacity of the parent sample and the intermediate amount of the N-doped sample. Fig. 4(c) shows the $C_p/T$ vs $T^2$ curve, with apparent anomaly jumps of around 2.7 K and 4.6 K, respectively, consistent with the $T_c$ in Table S2. The value of $\gamma$ and $\beta$ for α-$Mo_2C$ and $Mo_2C_{0.89}N_{0.11}$ are 5.618 mJ mol$^{-1}$ K$^{-2}$ and 0.089 mJ mol$^{-1}$ K$^{-4}$, 7.993 mJ mol$^{-1}$ K$^{-2}$ and 0.155 mJ mol$^{-1}$ K$^{-4}$, respectively. Fig. S6 shows the $C_{el}/T$ vs. $T$ curve under zero magnetic fields. The estimated $T_c = 2.58$ K and 4.51 K are also close to the $T_c$ of α-$Mo_2C$ and $Mo_2C_{0.89}N_{0.11}$ obtained from resistivity and magnetic susceptibility measurements. The calculated values of $\Delta C/\gamma T_c$ of 1.24 and 1.08 also demonstrate the presence of bulk superconductivity. Based on the above parameters, the Debye temperature can be obtained by $\Theta_D = \left(\frac{12\pi^4 nR}{5\beta}\right)^{1/3}$, where $n$ is the number of atoms per formula unit, and $R$ is the gas constant. The electron-phonon coupling constant $\lambda_{ep} = 0.66$ of $Mo_2C_{0.51}N_{0.49}$ can be calculated from the inverted McMillan formula[49]:

$\lambda_{ep} = \frac{1.04 + \mu^* \ln\left(\frac{\Theta_D}{1.45 T_c}\right)}{(1 - 1.62\mu^*) \ln\left(\frac{\Theta_D}{1.45 T_c}\right) - 1.04}$, using a typical Coulomb pseudopotential $\mu^*$ of 0.13[50-53]. The above parameters are summarized in Table 1, which shows that the nitrogen doping leads to the increase of electron-phonon coupling constant ($\lambda_{ep}$). Within the framework of BCS theory, $T_c$ depends on both the density of states at the Fermi level [$N(E_F)$] and the strength of electron-phonon coupling. Notably, when $x$ increases from 0 to 0.49, $T_c$ is enhanced more than double, and both $\gamma$ and $\lambda_{ep}$ increase. Therefore, the augmented of $T_c$ is likely due to the increase in $N(E_F)$ and the strengthening of electron-phonon coupling. Based on similar studies and theories[46,49], we can further analyze and speculate that the increase in $\lambda_{ep}$ can be attributed to phonon softening, which is consistent with the decrease in the Debye temperatures ($\Theta_D$). When $x = 0.11$ and $0.49$, the $\Theta_D$ drops to 335K and 377 K, respectively, indicating that phonon softening induced by nitrogen doping plays some role in enhancing the superconductivity of α-Mo$_2$C. However, since $\Theta_D$ does not change monotonically with $x$, this suggests that the effect of phonons may not be as crucial as the DOS.

We have plotted the electronic phase diagram of Mo$_2$C$_{1-x}$N$_x$ ($0 \leq x \leq 0.49$) to summarize the superconductivity of the N-doped α-Mo$_2$C system. Using the criterion that a 90 % fixed percentage of the normal-state resistivity is defined as the superconducting transition temperature of the sample, the curves obtained from the resistivity experimental results of $T_c^{onset}$ versus N doping concentration dependence are summarized. Fig. 5 shows that $T_c$ tends to increase with an increasing N content, reaching a maximum value of 7.11 K for MoC$_{0.51}$N$_{0.49}$. Subsequently, $T_c^{onset}$ decreases in the region of higher N doping values[3,23], demonstrating that N chemical doping can be used to adjust the superconducting properties of α-Mo$_2$C.

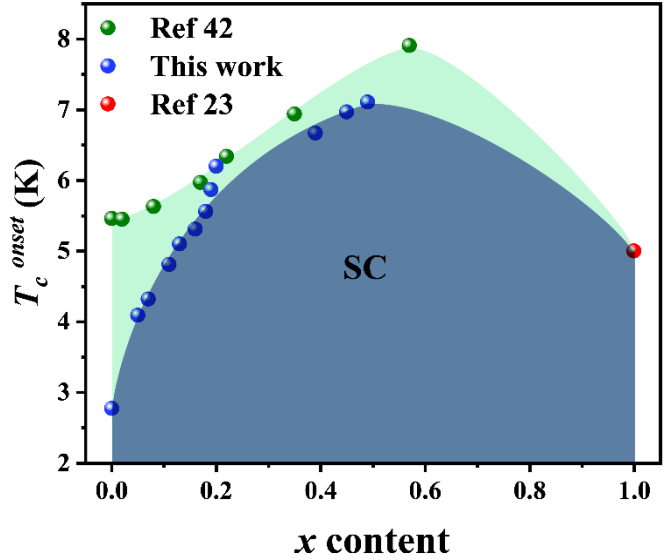

Fig. 5. The electronic phase diagram for Mo$_2$C$_{1-x}$N$_x$ vs N content.

Finally, first-principles calculations are employed for further understanding the increase of $T_c$ observed in the N-doped α-Mo$_2$C system. Our calculated electronic band structure [in Fig. 6(a)] and projected DOS [in Figs. 6(b), 6(c), and 6(d)] show that Mo$_2$C$_{0.8}$N$_{0.2}$ and Mo$_2$C$_{0.51}$N$_{0.49}$ are both metallic with a larger DOS at E$_F$ (6.920 and 7.887 states/eV, respectively) than that of α-Mo$_2$C (6.544 states/eV). As shown in the projected DOS, the electrons of Mo atoms have the dominant contribution to the DOS near E$_F$ and play an essential role in the superconductivity in these systems. We find that the $T_c$ of α-Mo$_2$C rises with the increase of N doping concertation, which can be explained by the increase of DOS at E$_F$ in the McMillan formula[54-56]. Our DFT calculations of DOS at E$_F$ for other compositions are shown in Fig. S7. Noteworthy, the electronic specific heat coefficients ($\gamma$) are proportional to the DOS value at E$_F$, consistent with the above analysis results of heat capacity data.

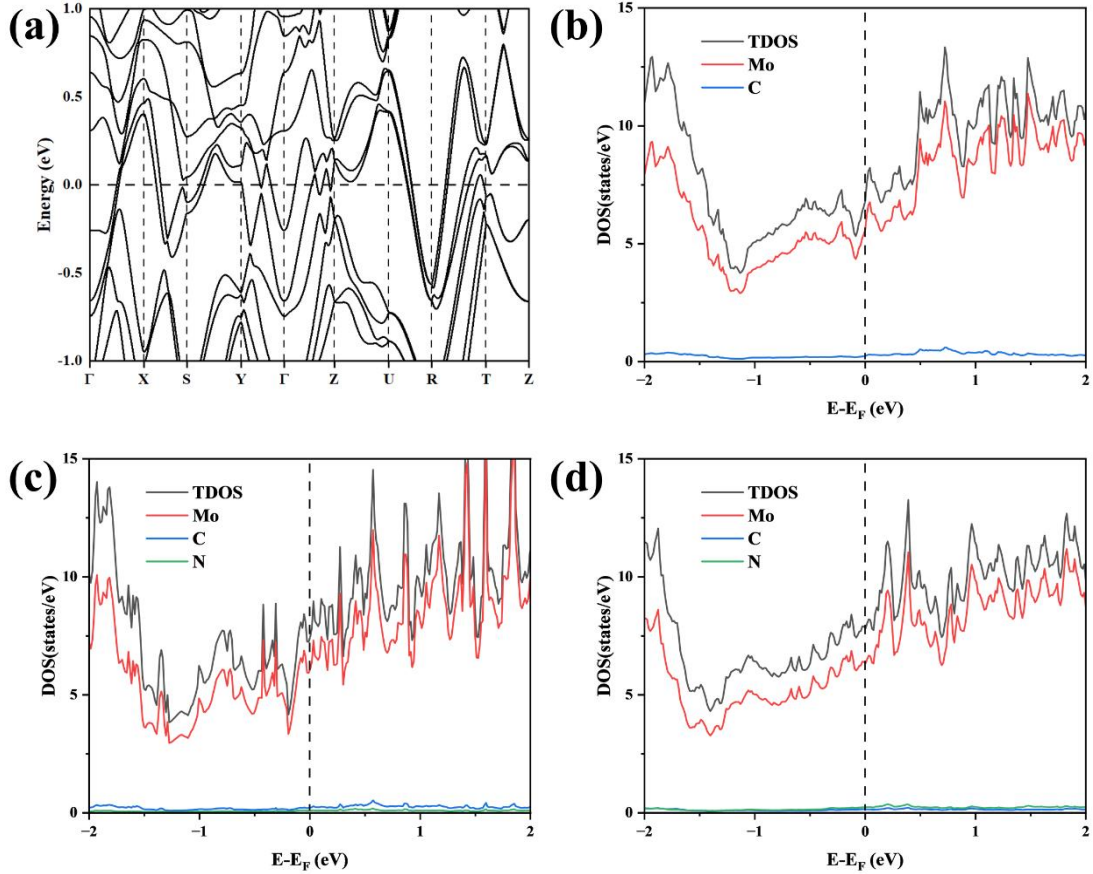

Fig. 6. (a) The electronic band structure of $Mo_2C_{0.51}N_{0.49}$ was calculated with spin-orbital coupling. The calculated projected density of state (DOS) of (b) α-$Mo_2C$, (c) $Mo_2C_{0.8}N_{0.2}$, and (d) $Mo_2C_{0.51}N_{0.49}$. TDOS is the total DOS of the system. The black dashed line indicates the Fermi level ($E_F$).

In conclusion, we propose a novel urea-assisted strategy for the preparation of $Mo_2C_{1-x}N_x$ ($0 \leq x \leq 0.49$) samples and systematically investigate how N doping influences the structure and superconducting properties of α-$Mo_2C$. The $Mo_2C_{1-x}N_x$ synthesized by this simple method exhibits a higher nitrogen content, and the N doping played a positive role in the increase of $T_c$. At the optimum doping of $Mo_2C_{0.51}N_{0.49}$, the maximum value of $T_c$ is 7.05 K, which is an increase of 4.37 K over the undoped sample. Our analysis suggests that the enhancement of superconductivity arises from the increased DOS at $E_F$ due to doping and the strengthened electron-phonon coupling resulting from doping-induced phonon softening. This work not only provides a new method for the preparation of transition metal carbonitrides but also expands the understanding of the superconductivity of N-doping transition metal carbides.

**Acknowledgments**

This work is supported by the National Natural Science Foundation of China (12274471, 11922415), Guangdong Basic and Applied Basic Research Foundation (2022A1515011168), the Key Research & Development Program of Guangdong Province, China (2019B110209003). The experiments

# Supporting Information

**A simple Urea approach to N-doped α-Mo$_2$C with enhanced superconductivity**

Longfu Li(李龙夫)[1], Lei Shi(石磊)[2], Lingyong Zeng(曾令勇)[1], Kuan Li(李宽)[1], Peifeng Yu(余沛峰)[1], Kangwang Wang(王康旺)[1], Chao Zhang(张超)[1], Rui Chen(陈睿)[1], Zaichen Xiang(项载琛)[1], Yunwei Zhang(张云蔚)[2], Huixia Luo(罗惠霞)[1,*]

[1]School of Materials Science and Engineering, State Key Laboratory of Optoelectronic Materials and Technologies, Guangdong Provincial Key Laboratory of Magnetoelectric Physics and Devices, Key Lab of Polymer Composite & Functional Materials, Sun Yat-Sen University, Guangzhou 510275, China

[2]School of Physics, Guangdong Provincial Key Laboratory of Magnetoelectric Physics and Devices, Sun Yat-Sen University, 510275 Guangzhou, China

China

*Corresponding author E-mail: *luohx7@mail.sysu.edu.cn* (H. Luo)

**Tab. S1** Determination of carbon and nitrogen content and Porosity in $Mo_2C_{1-x}N_x$.

| Sample | C [%] | N [%] | C/N ratio | Porosity |
|---|---|---|---|---|
| $Mo_2C$ | - | - | - | 35.04% |
| $Mo_2C_{0.95}N_{0.05}$ | 4.01 | 0.21 | 19.503 | 33.95% |
| $Mo_2C_{0.93}N_{0.07}$ | 5.29 | 0.43 | 12.309 | 34.05% |
| $Mo_2C_{0.89}N_{0.11}$ | 5.09 | 0.63 | 8.060 | 35.64% |
| $Mo_2C_{0.87}N_{0.13}$ | 4.65 | 0.69 | 6.699 | 34.52% |
| $Mo_2C_{0.84}N_{0.16}$ | 4.84 | 0.92 | 5.279 | 35.07% |
| $Mo_2C_{0.82}N_{0.18}$ | 4.59 | 1.03 | 4.447 | 36.33% |
| $Mo_2C_{0.81}N_{0.19}$ | 4.72 | 1.13 | 4.190 | 39.96% |
| $Mo_2C_{0.80}N_{0.20}$ | 4.97 | 1.23 | 4.045 | 35.55% |
| $Mo_2C_{0.61}N_{0.39}$ | 3.03 | 1.94 | 1.560 | 33.56% |
| $Mo_2C_{0.55}N_{0.45}$ | 2.37 | 1.96 | 1.209 | 33.98% |
| $Mo_2C_{0.51}N_{0.49}$ | 1.77 | 1.69 | 1.046 | 32.09% |

**Tab. S2** Superconducting parameters of $Mo_2C_{1-x}N_x$ ($0 \leq x \leq 0.49$) samples. The $\mu_0H_{c2}$ is obtained by fitting the 50 % criterion of normal state resistivity values using GL and WHH models. $\mu_0H^P$ is Pauli paramagnetic limit. $N(E_F)$ is the DOS at the Fermi level.

| Samples | $T_c$ (K) | $\mu_0H_{c2}$ (T) (WHH) | $\mu_0H_{c2}$ (T) (G-L) | $\mu_0H^P$ (T) | $N(E_F)$ (states/eV) | $\xi_{GL}(0)$ (nm) |
|---|---|---|---|---|---|---|
| $Mo_2C$ | 2.68 | 0.33 | 0.41 | 4.98 | 6.544 | 28.3 |
| $Mo_2C_{0.95}N_{0.05}$ | 3.93 | 0.54 | 0.64 | 7.31 | 6.598 | 22.6 |
| $Mo_2C_{0.93}N_{0.07}$ | 4.22 | 1.06 | 1.19 | 7.85 | | 16.6 |
| $Mo_2C_{0.89}N_{0.11}$ | 4.69 | 1.53 | 1.71 | 8.72 | 6.920 | 13.8 |
| $Mo_2C_{0.87}N_{0.13}$ | 5.06 | 3.65 | 3.98 | 9.41 | | 9.1 |
| $Mo_2C_{0.94}N_{0.16}$ | 5.25 | 3.85 | 4.31 | 9.76 | 7.028 | 8.7 |
| $Mo_2C_{0.82}N_{0.18}$ | 5.47 | 4.03 | 4.48 | 10.17 | | 8.5 |
| $Mo_2C_{0.81}N_{0.19}$ | 5.78 | 4.17 | 4.68 | 10.75 | | 8.4 |
| $Mo_2C_{0.80}N_{0.20}$ | 6.01 | 4.10 | 4.67 | 11.17 | 7.246 | 8.4 |
| $Mo_2C_{0.61}N_{0.39}$ | 6.59 | 5.67 | 6.12 | 12.26 | 7.538 | 7.3 |
| $Mo_2C_{0.55}N_{0.45}$ | 6.88 | 6.54 | 7.02 | 12.79 | | 6.8 |
| $Mo_2C_{0.51}N_{0.49}$ | 7.05 | 5.75 | 6.16 | 13.11 | 7.887 | 7.2 |

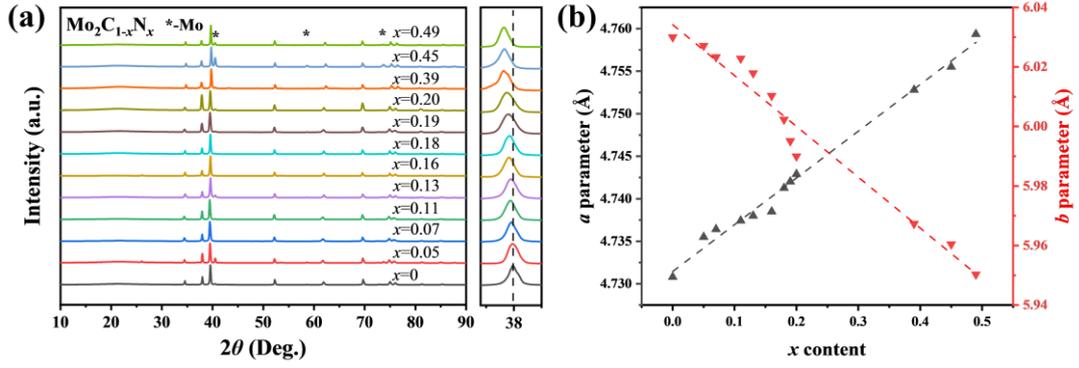

**Fig. S1** (a) PXRD patterns for Mo$_2$C$_{1-x}$N$_x$ (0 ≤ $x$ ≤ 0.49) samples. The peak in the representative lattice plane (200) was enlarged on the right. (b) The trends in lattice parameters as the change of the N doping amount.

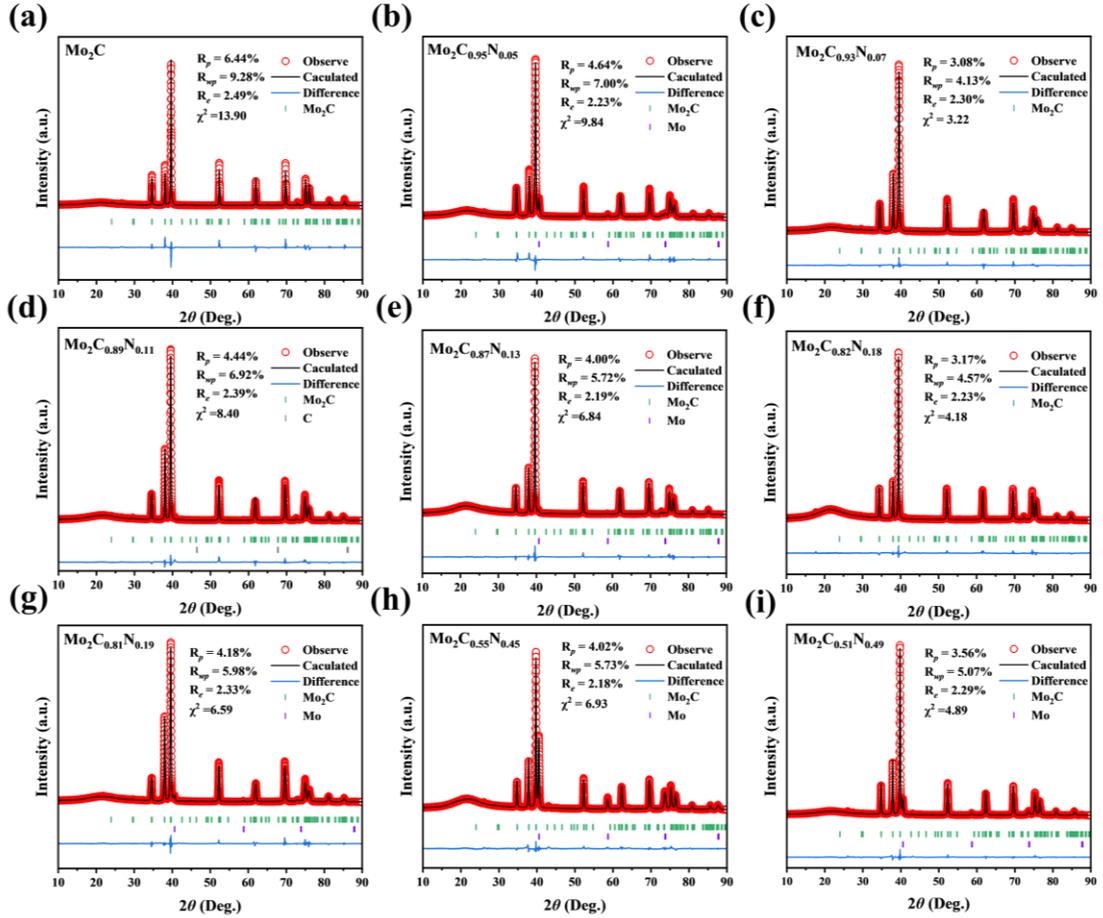

**Fig. S2** The XRD result after Rietveld refinement for Mo$_2$C$_{1-x}$N$_x$ (0 ≤ $x$ ≤ 0.49).

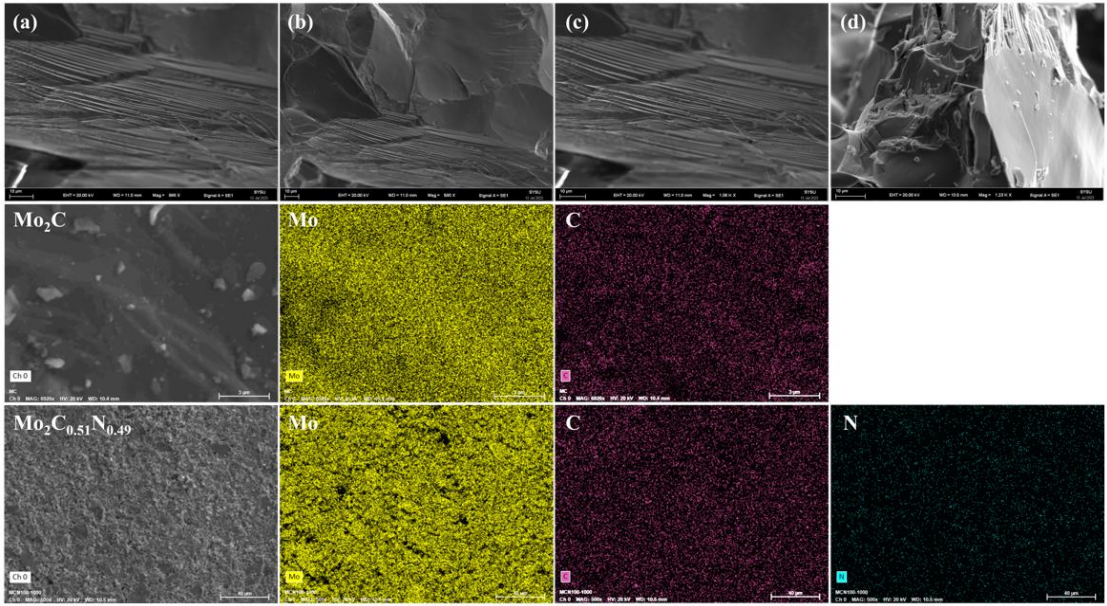

**Fig. S3** (a), (b), (c), and (d) show the SEM of $Mo_2C$ and EDS mappings of $Mo_2C$ and $Mo_2C_{0.51}N_{0.49}$.

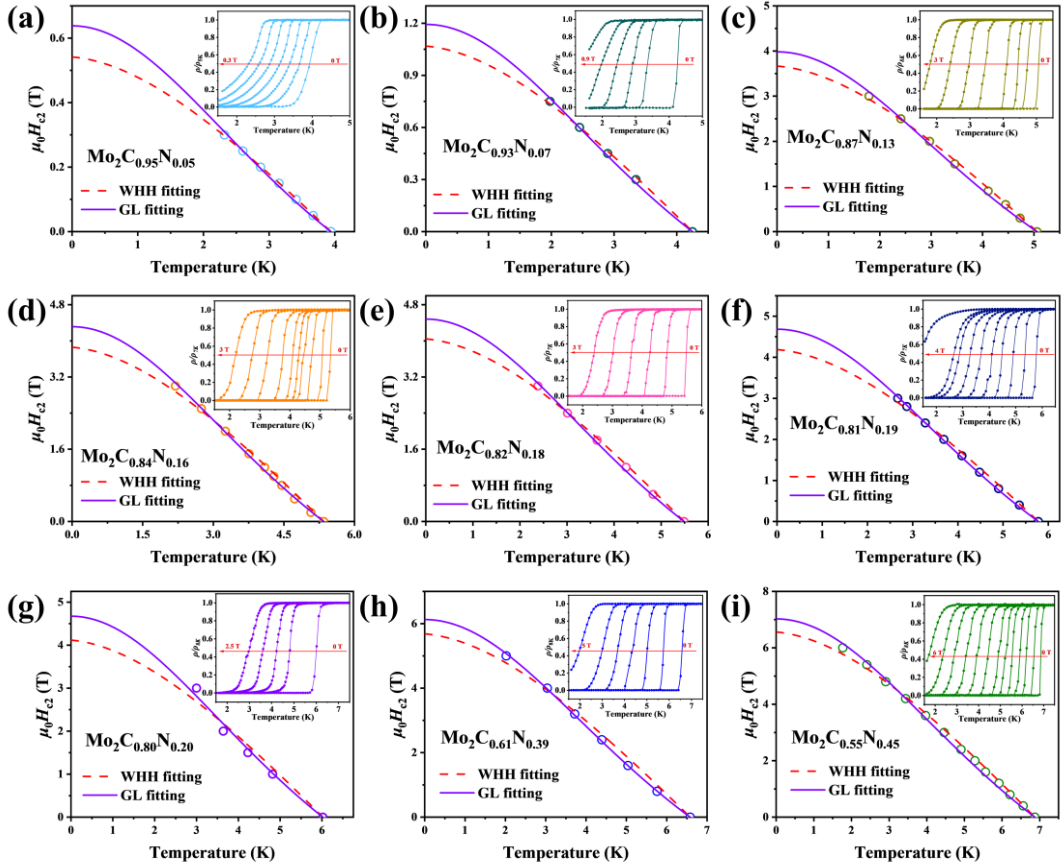

**Fig. S4** The upper critical fields of $Mo_2C_{1-x}N_x$ ($0.05 \leq x \leq 0.45$) samples. The top right inset shows the low-temperature resistivity at different applied fields.

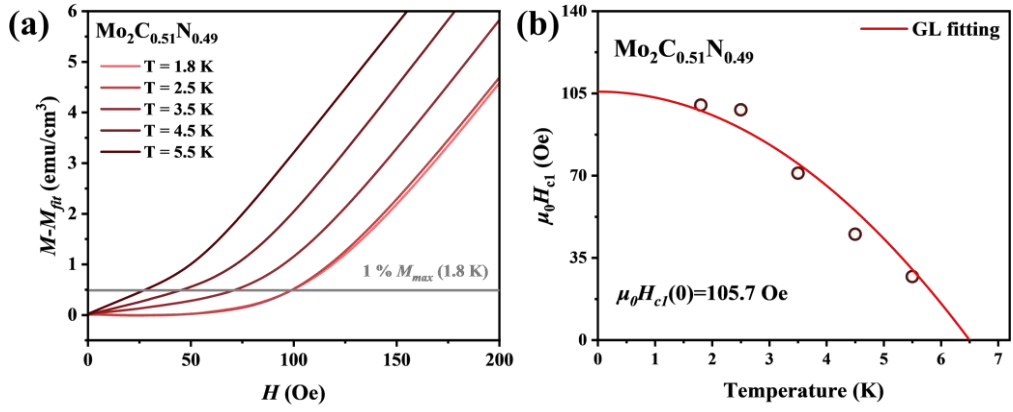

**Fig. S5** (a) The value of M-M$_{fit}$ at 0-200 Oe under several temperatures. (b) The $\mu_0H_{c1}-T$ phase diagram for Mo$_2$C$_{0.51}$N$_{0.49}$.

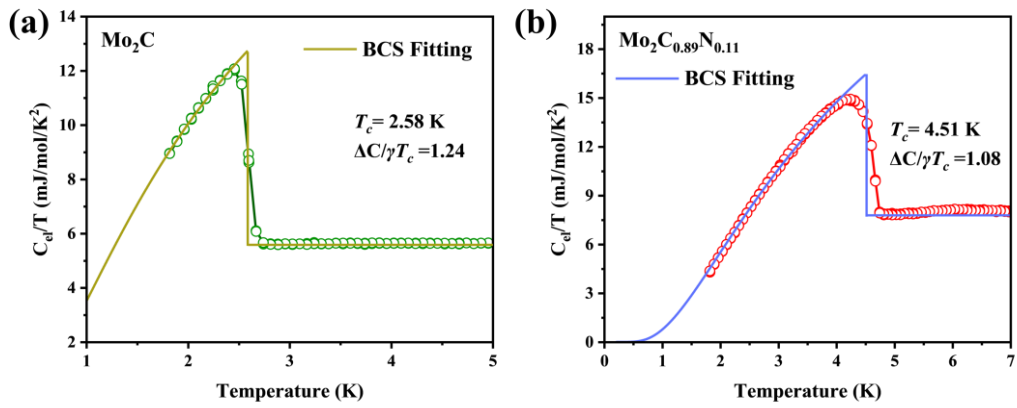

**Fig. S6** The temperature dependence of the electronic contribution to the specific heat for Mo$_2$C and Mo$_2$C$_{0.89}$N$_{0.11}$.

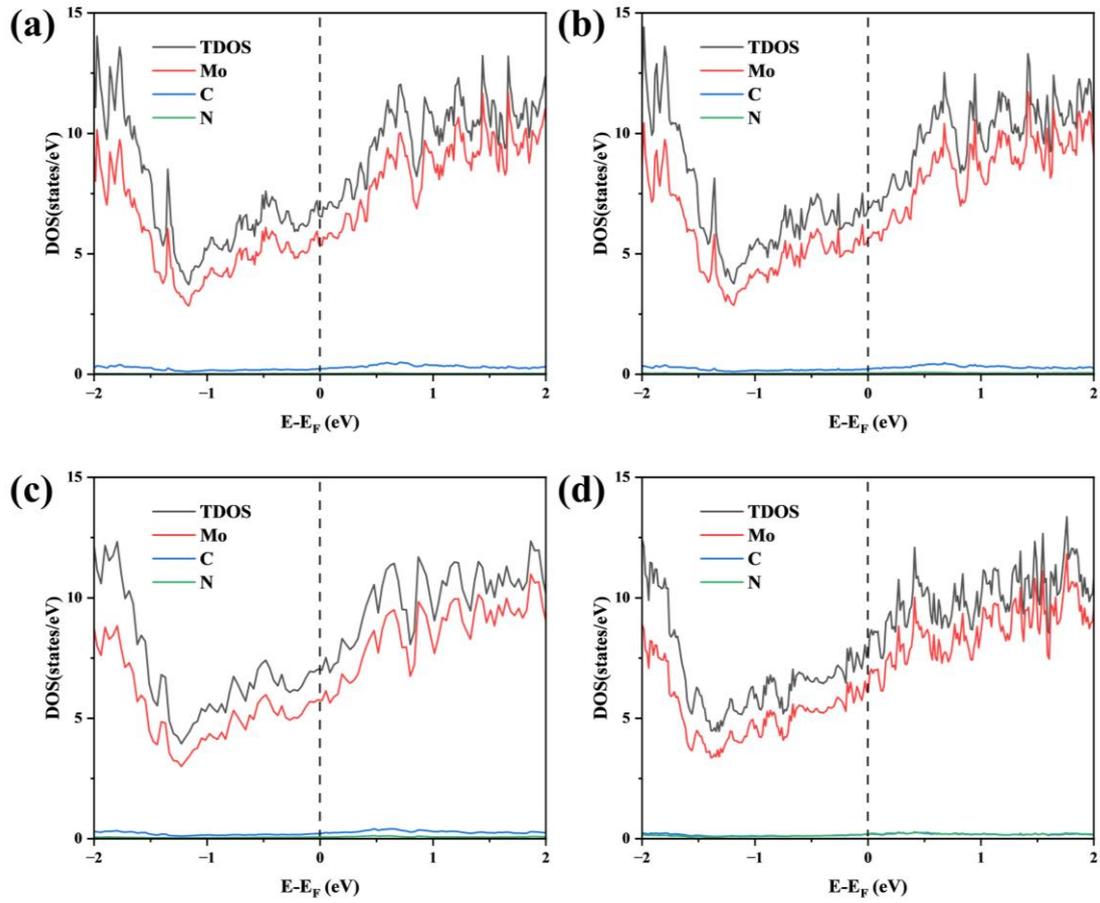

**Fig. S7** The calculated projected density of state (DOS) of (a) $Mo_2C_{0.95}N_{0.05}$, (b) $Mo_2C_{0.89}N_{0.11}$, (c) $Mo_2C_{0.84}N_{0.16}$ and (d) $Mo_2C_{0.61}N_{0.39}$. TDOS is the total density of state of the system. The black dashed line indicates the Fermi level ($E_F$).